\documentclass[preprint2]{aastex}

\textheight=\baselinestretch\textheight
\ifdim\textheight>25.2cm\textheight=25.0cm\fi

\topskip\baselineskip
\maxdepth\baselineskip
\let\tighten=\relax
\let\tightenlines=\tighten

\newcommand{\mic}{\mbox{$\mu{\rm m}$}}

\newcommand{\xten}[1]{\mbox{$\times 10^{#1}$}}

\newcommand{\kms}{\mbox{km\,s$^{-1}$}}
\newcommand{\degs}{\mbox{$^{\rm o}$}}
\newcommand{\mdot}{\mbox{$\stackrel{.}{M}$}}

\newcommand{\uchii}{\mbox{UCH\,{\scriptsize II}}}
\newcommand{\hii}{\mbox{H\,{\scriptsize II}}}

\slugcomment{To appear in PASP.}

\shorttitle{The CORNISH Survey}
\shortauthors{Hoare}

\begin{document}

\title{The Co-ordinated Radio and Infrared Survey for High Mass Star
  Formation (The CORNISH Survey) - I. Survey Design}

\author{M.~G.~Hoare\altaffilmark{1}, 
C.~R.~Purcell\altaffilmark{1,2,21},  
E.~B.~Churchwell\altaffilmark{3}, 
P.~Diamond\altaffilmark{2,17}, 
W.~D.~Cotton\altaffilmark{4}, 
C.~J.~Chandler\altaffilmark{5}, 
S.~Smethurst\altaffilmark{2}, 
S.~E.~Kurtz\altaffilmark{6}, 
L.~G.~Mundy\altaffilmark{7}, 
S.~M.~Dougherty\altaffilmark{8}, 
R.~P.~Fender\altaffilmark{9},
G.~A.~Fuller\altaffilmark{2}, 
J.~M.~Jackson\altaffilmark{10}, 
S.~T.~Garrington\altaffilmark{2}, 
T.~R.~Gledhill\altaffilmark{11}, 
P.~F.~Goldsmith\altaffilmark{12},
S.~L.~Lumsden\altaffilmark{1}, 
J.~Mart\'{i}\altaffilmark{16}, 
T.~J.~T.~Moore\altaffilmark{13}, 
T.~W.~B.~Muxlow\altaffilmark{2}, 
R.~D.~Oudmaijer\altaffilmark{1}, 
J.~D.~Pandian\altaffilmark{14}, 
J.~M.~Paredes\altaffilmark{15}
D.~S.~Shepherd\altaffilmark{5,19}, 
R.~E.~Spencer\altaffilmark{2}, 
M.~A.~Thompson\altaffilmark{11}, 
G.~Umana\altaffilmark{18}, 
J.~S.~Urquhart\altaffilmark{1,17,20}
A.~A.~Zijlstra\altaffilmark{2}}

\altaffiltext{1}{School of Physics and Astronomy, University of Leeds, Leeds, LS2 9JT, UK} 
\altaffiltext{2}{Jodrell Bank Centre for Astrophysics, The University of Manchester
Alan Turing Building, Manchester, M13 9PL, UK} 
\altaffiltext{3}{The University of Wisconsin, Department of Astronomy, 475 North Charter Street
Madison, WI 53706, USA}
\altaffiltext{4}{National Radio Astronomy Observatory, 520 Edgemont Road, Charlottesville, VA 22903-2475, USA}
\altaffiltext{5}{National Radio Astronomy Observatory, P.O. Box O, Socorro, NM 87801-0387, USA}
\altaffiltext{6}{Centro de Radioastronom\'{i}a, Universidad Nacinal
  Aut\'{o}noma de M\'{e}xio - Morelia, Apartado Postal 3-72, C.P. 58090 Morelia, Michoacan, Mexico}
\altaffiltext{7}{Department of Astronomy, University of Maryland College Park, MD 20742-2421, USA}
\altaffiltext{8}{National Research Council of Canada, Herzberg Institute for Astrophysics, Dominion Radio Astrophysical Observatory, PO Box 248, Penticton, British Columbia V2A 6J9, Canada}
\altaffiltext{9}{School of Physics and Astronomy, University of Southampton, Southampton SO17 1BJ}
\altaffiltext{10}{Astronomy Department, Boston University, 725 Commonwealth Avenue, Boston, MA 02215, USA}
\altaffiltext{11}{Science and Technology Research Institute, University of Hertfordshire, College Lane, Hatfield AL10 9AB, UK}
\altaffiltext{12}{Jet Propulsion Laboratory, 4800 Oak Grove Drive, Pasadena, California 91109, USA}
\altaffiltext{13}{Astrophysics Research Institute, Liverpool John Moores University, Twelve Quays House, Egerton Wharf, Birkenhead CH41 1LD, UK}
\altaffiltext{14}{Indian Institute
of Space Science and Technology, Valiamala, Trivandrum 695547, India}
\altaffiltext{15}{Departament d'Astronomia i Meteorologia and Institut de Ci\`{e}ncies del Cosmos (ICC), Universitat de Barcelona (UB/IEEC), Mart\'{i} i Franqu\`{e}s 1, 08028 Barcelona, Spain}
\altaffiltext{16}{Departamento de F\'{i}sica, EPS, Universidad de Ja\'{e}n, Campus Las Lagunillas, Edif. A3, 23071 Ja\'{e}n, Spain}	
\altaffiltext{17}{CSIRO Australia Telescope National Facility, PO Box 76, Epping NSW 1710, Australia}
\altaffiltext{18}{Osservatorio Astrofisico di Catania, Via S.Sofia 78, 95123 Catania, Italy}
\altaffiltext{19}{Square Kilometer Array—Africa, 3rd Floor, The Park, Park Road, Pinelands, 7405, South Africa}
\altaffiltext{20}{Max Planck Institute for Radio Astronomy, P.O. Box 20 24, 53010 Bonn, Germany}
\altaffiltext{21}{Sydney Institute for Astronomy, School of Physics, The University of Sydney, NSW 2006,
Australia}

\email{m.g.hoare@leeds.ac.uk}

\begin{abstract}
We describe the motivation, design and implementation of the CORNISH
survey, an arcsecond resolution radio continuum survey of the inner
Galactic plane at 5\,GHz using the Karl G. Jansky Very Large Array (VLA). It is a blind survey
co-ordinated with the northern {\it Spitzer} GLIMPSE I region covering
10\degs$<l<$65\degs\ and $|b|<$1\degs\ at similar resolution. 
We discuss in detail the strategy that we employed to control the
shape of the synthesised beam across this survey that covers a wide
range of fairly low declinations. Two snapshots separated by 4 hours
in hour angle kept the beam elongation to less that 1.5 over 75\,\% of
the survey area and less than 2 over 98\,\% of the survey. The
prime scientific motivation is to provide an unbiased survey for ultra-compact $\hii$~regions to study this key phase in massive star formation.  A
sensitivity around 2\,mJy will allow the automatic distinction between
radio loud and quiet mid-IR sources found in the {\it Spitzer} surveys. This
survey has many legacy applications beyond star formation including
evolved stars, active stars and binaries, and extragalactic
sources. The CORNISH survey for compact ionized sources complements
other Galactic plane surveys that target diffuse and non-thermal sources as
well as atomic and molecular phases to build up a complete picture of
the ISM in the Galaxy. 
\end{abstract}



\section{Introduction}

The study of star formation, stellar populations and Galactic
structure is currently being transformed by a new generation of
Galactic plane surveys. These surveys have much higher resolution and
sensitivity than previous surveys and are at infrared and longer wavelengths
that can penetrate the high extinction of the Galactic mid-plane. They
also cover sufficiently wide areas of the plane to provide samples that
are large enough to be representative and statistically robust for
rare and short-lived phases of evolution.  Blind surveys that cover
contiguous, wide areas also allow the study of phenomena and trends
across a large range of scales.  A
wide wavelength coverage is required to distinguish and characterise the
different populations found at both the early and later phases of
stellar evolution and so it is also important to ensure that the same
wide areas are covered by all the relevant wavelengths. Here we
describe the design and implementation of a new radio continuum survey of the Galactic
plane and its role in the context of the other multi-wavelength
surveys.

\subsection{A New Generation of Galactic Plane Surveys}

Advances in various wide-field technologies have led to a new
generation of Galactic plane surveys.  Those with relevance here, and
to star formation in particular, are listed in
Table~\ref{surveys}. Leading the way is the GLIMPSE I (Galactic Legacy 
Infrared Mid-Plane Survey Extraordinaire) survey (Benjamin et
al. 2003; Churchwell et al. 2009). This {\it Spitzer} IRAC Legacy Programme
covered the inner galaxy ($10\degs<l<65\degs$ and
$-65\degs<l<-10\degs$, $|b|<1\degs$) at 3.6\,\mic, 4.5\,\mic, 5.8\,\mic\ and
8.0\,\mic. GLIMPSE is two orders of magnitude more sensitive (limit at
L-band (3.6\,\mic)\,$\approx14-15$ magnitude although usually confusion
limited) and has ten times higher spatial resolution ($1.4-1.9\arcsec$)
than any previous mid-IR survey. It has catalogued over 49 million
sources in a wavelength regime which preferentially selects sources
with hot circumstellar dust emission such as young and evolved stars
(e.g., Robitaille et al. 2008). More embedded and cooler sources are
the subject of the MIPSGAL survey with {\it Spitzer} (Carey et al. 2009) and
the PACS element of the Hi-GAL survey with Herschel (Molinari et
al. 2010).

\begin{deluxetable}{@{}l@{}c@{}c@{}c@{~}c@{~}c@{~}l@{}}
\tablecaption{Galactic plane surveys showing the context of the CORNISH
  survey. \label{surveys}}
\tablewidth{0pt}
\tablehead{
\colhead{Survey} & \colhead{Wavelength} & \colhead{Beam (\arcsec)} &
\colhead{$l$ Coverage}& \colhead{$b$ Coverage} &
\colhead{Probe} &
\colhead{Reference}}
\startdata
IPHAS & H$\alpha$  & 1.7 & $30$\degs $<l<210$\degs & $|b|<5$\degs &
Nebulae \& stars  & Drew et al. (2005) \\
UKIDSS & JHK & 0.8 &  $-2$\degs $<l<230$\degs & $|b|<1$\degs &  Stars,
Nebulae & Lucas et al. (2008) \\
VVV & ZYJHK & 0.8 & $-65$\degs $<l<10$\degs & $|b|<2$\degs & `` & Minniti
et al. (2010)  \\
GLIMPSE & 4-8\,\mic & 2 &  $-65$\degs $<l<65$\degs  &$|b|<1$\degs &
Stars, Hot Dust & Churchwell et al. (2009) \\
MSX & 8-21\,\mic & 18 & All &$|b|<5$\degs & Warm Dust & Price et
al. (2001) \\
MIPSGAL & 24,70\,\mic & 6, 20 & $-65$\degs $<l<65$\degs &$|b|<1$\degs &
`` & Carey et al. (2009) \\
AKARI & 50-200\,\mic & 30-50 & All sky & & Cool Dust & White et
al. (2009) \\
Hi-GAL & 70-500\,\mic & 10-34 &  All  & $|b|<1$\degs\tablenotemark{a}
& `` & Molinari et al. (2010) \\
JPS   & 450,850\,\mic & 8-14 & $10$\degs $<l<60$\degs &$|b|<1$\degs & ``
& Moore et al. (2005) \\
ATLASGAL & 850\,\mic & 19 &$-60$\degs $<l<60$\degs &$|b|<1.5$\degs & `` &
Schuller et al. (2009) \\
BOLOCAM & 1100\,\mic & 33 & $-10$\degs $<l<90$\degs & $|b|<0.5$\degs& ``
&
Aguirre et al. (2010) \\
GRS  & $^{13}$CO 1-0 & 46 & $18$\degs $<l<56$\degs &$|b|<1$\degs&
Molecular Gas & Jackson et al. (2006) \\
MMB & 6.7\,GHz & 192\tablenotemark{b} & $-180$\degs $<l<60$\degs &$|b|<2$\degs &
Methanol Masers & Green et al. (2009) \\
HOPS & 22\,GHz & 132\tablenotemark{b} & $-180$\degs $<l<60$\degs &$|b|<2$\degs & Water
Masers & Walsh et al. (2011) \\
OH & 1.6\,GHz & $\sim$10 &  $-45$\degs $<l<45$\degs & $|b|<3$\degs & Hydroxyl
Masers & Sevenster et al. (2001) \\
CORNISH & 6 cm & 1.5 &  $10$\degs $<l<65$\degs &$|b|<1$\degs &Compact
Ionized Gas & This work \\
S/V/CGPS & 21 cm & 60 &  $-107$\degs $<l<147$\degs & $|b|<1.3$\degs
&Atomic Gas & Stil et al. (2006) \\
MAGPIS & 20 cm & 5 &  $5$\degs $<l<48$\degs & $|b|<0.8$\degs &Diffuse
Ionized Gas & Helfand et al. (2006) \\
MGPS-2  & 35 cm & 45 &  $-115$\degs $<l<0$\degs &$|b|<10$\degs & `` &
Murphy et al. (2007) \\
\enddata
\tablenotetext{a}{Follows warp} 
\tablenotetext{b}{Maser positions are accurate to about
  0.1\arcsec\ after interferometric follow-up.} 
\end{deluxetable}

An even deeper probe of the general stellar population is becoming
available via the deep near-IR Galactic Plane Survey (GPS) that is
part of the UK IR Deep Sky Surveys (UKIDSS) programme (Lucas et
al. 2008).  This survey is about 3 magnitudes more sensitive and two
to three times better resolution than the 2MASS all sky near-IR survey
(Skrutskie et al. 2006) and will detect upwards of a billion sources.
The addition of near-IR data, where photospheric and scattered light
contributions dominate, allows a much better separation and
characterisation of the general stellar population, than from mid-IR
colours alone. Near-IR surveys are also being used to map the
extinction due to molecular clouds (e.g., Rowles \& Froebrich
2009). The optical H$\alpha$ survey IPHAS (Drew et al. 2005) is
providing both stellar characterisation, 3D extinction maps and
tracing nebular emission in less obscured regions of the northern
plane with the southern VPHAS+ survey (see www.vphasplus.org)
scheduled to do the same in the south.

To place the stellar populations in a Galactic context, commensurate
studies of the various components of the ISM are required. For the
study of star formation in particular, the molecular component is
provided in part by the BU-FCRAO $^{13}$CO\,1-0 Galactic Ring Survey
(Jackson et al. 2006), which covers over half of the northern GLIMPSE
area with extensions over much of the northern mid-plane (Mottram \&
Brunt 2010). The mostly optically thin $^{13}$CO\,1-0 data traces the
distribution and dynamics of the cold molecular gas. The cool dust
in these molecular clouds can also be mapped via its sub-millimetre
continuum emission. This is currently being achieved with the ATLASGAL
survey (Schuller et al. 2009) at 870\,\mic\ and BOLOCAM survey at
1100\,\mic\ (Aguirre et al. 2010), whilst the forthcoming JCMT
Galactic Plane survey (JPS, www.jach.hawaii.edu/JCMT/surveys/) will
map the continuum at 450 and 850\,\mic\ with higher sensitivity and
resolution. Together with the SPIRE element of the Hi-GAL survey
with the Herschel satellite (Molinari et al. 2010) this will provide
temperature and dust emissivity information across the GLIMPSE region.

The more widely distributed atomic hydrogen component is covered by
the International Galactic Plane Survey at about 1\arcmin\ resolution
(www.ras.ucalgary.ca/IGPS/). This is
made up of the VLA Galactic Plane Survey (VGPS, Stil et al. 2006),
Southern Galactic Plane Survey (SGPS; McClure-Griffiths et al. 2005)
and Canadian Galactic Plane Survey (Taylor et al. 2003). A sub-section
of the plane is covered by the much deeper GALFA survey of
H\,{\scriptsize I} at
4\arcmin\ resolution using the Arecibo telescope (Peek et al. 2011).
The H\,{\scriptsize I} component traces where most of the mass of
interstellar gas is 
located and its velocity structure.  21\,cm absorption lines can also
help to solve the near/far distance ambiguity in relation to the kinematic
distances derived towards the embedded $\hii$~regions (Sewilo et
al. 2004; Fish et al. 2003)  and dense, molecular clouds (Jackson et
al. 2002; Gibson et al. 2005; Busfield et al. 2006; Roman-Duval et
al. 2009). This combination of atomic and molecular kinematic data can
therefore provide the 3D Galactic setting of the neutral gas.

The one major component missing from the multi-wavelength surveys of
the Galaxy is that of the ionized gas. No existing radio continuum
survey has sufficient resolution, depth or coverage to provide the
data that would complete the picture of the Galaxy. Ionized gas arises
in nebulae and stellar winds around hot stars and is crucial to
understanding key phases of the early and late evolution of high and
intermediate mass stars.  Dense, photo-ionized regions around young and
evolved stars are often optically thick with a turnover frequency of
around a few GHz, where the radio continuum spectral index $\alpha$
($S_{\nu}\propto\nu^{\alpha}$) changes from the optically thick value
of +2.0 to the optically thin slope of $-0.1$.
Ionized stellar winds have a positive spectral index $\alpha\sim+0.6$,
and so it is also much more efficient to carry out systematic 
searches for these sources at high frequencies. Furthermore, the
resolution should be comparable to the arcsecond resolution of the IR
surveys to allow IR counterparts to be uniquely identified. It was to
fill this need for a high frequency, high resolution radio continuum
survey covering the area of the {\it Spitzer} survey that the concept
of the Co-Ordinated Radio `N' Infrared Survey for High-mass star
formation (CORNISH\footnote{This acronym derives from the Celtic
  origins of Hoare and Diamond.}) survey was conceived.

\subsection{Previous Radio Surveys}

Many single-dish surveys of the Galactic plane have been carried out
at 5\,GHz with a resolution of a few arcminutes that is totally
inadequate to complement the new generation of IR surveys
(e.g., Altenhoff et al. 1978; Haynes et al. 1978; Gregory \& Condon
1991; Griffith et al. 1994). Most higher resolution interferometric
surveys have been carried out at the relatively low frequency of 1.4
GHz (Zoonematkermani et al. 1990; Becker et al. 1990; Condon et
al. 1998; Giveon et al. 2005).  Although the multi-configuration
MAGPIS survey (Helfand et al. 2006) and MGPS-2 survey (Murphy et
al. 2007) are good for tracing the optically thin thermal
emission from extended, evolved $\hii$~regions and non-thermal
supernova remnants, they are not appropriate for the study of dense,
thermal sources due to the unfavourable spectral index and
compactness.  Becker et al. (1994), with infills by Giveon et
al. (2005) and White et al. (2005) have surveyed part of the inner
galaxy ($-10\degs<l<42\degs$, $|b|<0.4\degs$) at 5\,GHz with the VLA in
C configuration giving a resolution of
4\arcsec$\times$9\arcsec. However, these surveys cover only one fifth
of the northern GLIMPSE region and the spatial resolution is poorer than
that of {\it Spitzer} IRAC, which causes problems with source
identification. At this resolution the vast majority of their sources
are unresolved and thus we have no morphological information. In the
southern hemisphere the AT20G survey at 20\,GHz and
$\sim$30\arcsec\ resolution (Murphy et al. 2010) that was primarily designed
to study the extragalactic radio source population is detecting many 
$\hii$~regions, but again not at sufficient depth or resolution to address
the science opened up by current surveys of the Galactic plane at infrared
wavelengths.

\section{Scientific Motivation}
\label{science}

In this section we describe the various science topics that are
addressed by the CORNISH survey. These cover the full range of
galactic science from the formation and evolution of stars through
to background extra-galactic sources. 

\subsection{High-Mass Star Formation}

Our main science driver is the study of the formation of massive
stars. These stars impart large amounts of UV radiation, wind energy,
strong shocks and chemically enriched material into the ISM during their
lifetime. This is particularly true in starbursts when they form in
dense concentrations (Leitherer 1999).  However, relatively little is
known about the physics that controls the formation of stars more
massive than about 8\,M$_{\odot}$ (10$^{3}$\,L$_{\odot}$) and hence the
upper initial mass function (IMF). The fast collapse and contraction
time scales, and extreme conditions resulting from increased
turbulence and radiation pressure makes the theoretical treatment of
their formation more challenging than for solar-type stars
(e.g., Behrend \& Maeder 2001; Yorke \& Sonnhalter 2002; McKee \& Tan
2003; Bonnell \& Bate 2006).  Their rarity and predilection to form in
dense clusters makes observational studies more challenging and high
resolution a pre-requisite (e.g., Beuther et al. 2002).

Current numerical simulations indicate that accretion via a disc can
continue despite the strong radiation pressure (Krumholz et al. 2009;
Kuiper et al. 2010)
whilst sub-millimetre interferometry reveals evidence of flattened,
disc-like structures (e.g., Patel et al. 2005; Torrelles et al. 2007) around
massive young stellar objects (MYSOs).  Although these deeply embedded
IR objects are already very luminous, they have yet to ionize the
surrounding ISM, most likely due to ongoing accretion keeping them
swollen and cool (Hoare \& Franco 2007; Hosokawa \& Omukai 2009;
Davies et al. 2011).
MYSOs drive highly collimated ionized jets (Mart\'{i} et
al. 1998; Curiel et al. 2006) and equatorial disc winds (Hoare et
al. 1994; Hoare 2006; Gibb \& Hoare
2007) which are manifest as compact thermal radio emission (a few
mJy for nearby examples) with a spectral index close to +0.6.  What
drives these jets and winds at speeds of a few hundreds of \kms\
(e.g., Bunn et al. 1995) is not known. A possible scenario is that the
radio jets are accelerated and collimated by an MHD mechanism as in
low-mass YSOs early on, and these give way to less-collimated radiation
pressure dominated flows (Drew et al. 1998) at later stages, akin to
the scheme outlined by Beuther \& Shepherd (2005).

The strong radio emission from $\hii$~regions, formed once the 
star has contracted down to the main sequence and heated up, is a
very useful probe of massive star formation (Hoare et al. 2007). It is
unaffected by extinction and can thus be used to measure the formation
rate of OB stars right across a galaxy (e.g., Condon, Cotton \&
Broderick 2002). The radio brightness of $\hii$~regions can be used to
infer the flux of Lyman continuum photons from the exciting star or
cluster and hence constrain the spectral type and mass of the
star(s). $\hii$ regions provide one of the clearest signatures of spiral
structure which can be related to the global distribution of molecular
and atomic hydrogen from complementary surveys. This is
particularly important in the context of the emerging picture of our
Galaxy as having two stellar arms linked to the central bar whilst
having four gaseous star forming arms (Benjamin 2009).

\begin{figure}
\centering
\includegraphics[width=8.5cm]{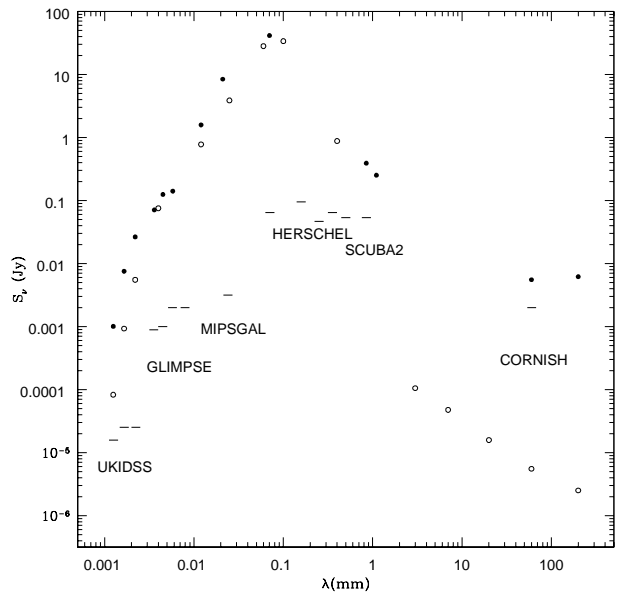}
\caption{Spectral energy distribution from the near-IR to the radio of
  an example $\uchii$~region (G042.4343$-$00.2597,
  L\,=\,1.8\xten{4}\,L$_\sun$, d\,=\,5.0\,kpc, solid circles) and MYSO
  (S140 IRS 1, L\,=\,0.85\xten{4}\,L$_\sun$, d\,=\,0.76 kpc, open
  circles). Sources of the IR data for the $\hii$~region 
  can be found in Mottram et al (2011) and the radio data come from
  Wood \& Churchwell (1989) and Garwood et al. (1988). For S140 IRS 1
  the IR data points are discussed in Maud et al. (2012) and the radio
  points are from Gibb \& Hoare (1997) and Schwartz (1989). The 
  flux densities from these objects have been reduced
  to place them at 15 kpc to reflect
  early B stars on the far side of the galaxy. The 2\,mJy (5$\sigma$)
  sensitivity limit for the CORNISH survey is marked along with the
  5$\sigma$ limits for the UKIDSS, GLIMPSE, MIPSGAL, Herschel and JCMT
  surveys. Note that the sensitivities are well matched to the SEDs of
  our main targets and that the radio data are vital for
  distinguishing between the radio-loud $\uchii$s and radio-quiet
  MYSOs. \label{cornishfig}}
\end{figure}

High resolution studies of Galactic objects have revealed both large
numbers of ultra-compact objects ($\uchii$s) and a preponderance of
cometary morphologies (Wood \& Churchwell 1989; Kurtz et al. 1994;
Walsh et al. 1998). These studies spawned a re-examination of the
dynamical models of $\hii$~regions (Van Buren \& Mac Low 1992; De Pree,
Rodr\'{i}guez \& Goss 1995; Williams, Dyson \& Redman 1996). The
latest cometary models (Arthur \& Hoare 2006) require the young OB
star to be located in a density gradient to explain the cometary
shape, together with a strong stellar wind to give the limb-brightened
appearance and an element of proper motion up the density gradient to
reconcile the ionized and molecular velocity structure. Such a picture
is consistent with one where the formation of these massive
stars was triggered by the passage of an external shock-wave through
the molecular cloud.

There are many instances where the expansion of $\hii$~regions appears
to be triggering formation of new episodes of massive star formation
(e.g., Deharveng et al. 2005; Urquhart, Morgan \& Thompson 2009; Moore
et al. 2007; Zavagno et al. 2010; Thompson et al. 2012).  This is
readily apparent in the 8\,\mic\ images from the GLIMPSE survey, where
the strong Polycyclic Aromatic Hydrocarbon (PAH) emission delineates
the Photo-Dissociation Region (PDR) at the interface between the
ionized region and the molecular cloud. Furthermore the
24\,\mic\ images from MIPSGAL show up the warm dust heated by
L$\alpha$ within the ionized zone and correlates well with the radio
free-free emission (e.g., Watson et al. 2008).  Younger phases such as
mid-IR bright MYSOs or maser sources are often seen just ahead of the
ionization front sometimes within IR dark regions (Hoare et al. 2007).
What fraction of massive star formation occurs in this way, as opposed
to spontaneous collapse is unclear. Conversely, the action of
outflows, expanding $\hii$~regions and stellar winds can also disperse
the remaining molecular cloud and quench further star formation (e.g.,
Franco, Shore \& Tenorio-Tagle 1994). How these opposing feedback
mechanisms affect the resultant star formation rate and efficiency is
unknown.

Many aspects of massive star formation studies suffer from the lack of
a large and unbiased sample of both $\uchii$s and MYSOs.  Samples based
on the IRAS Point Source Catalogue (e.g., Kurtz et al. 1994; Molinari
et al 1996; Sridharan et al. 2002) are heavily biased towards bright,
isolated sources due to the poor spatial resolution of IRAS, which
resulted in confusion in the densest regions of the plane, precisely where
massive stars form.  The situation for mid-IR bright MYSOs
has been alleviated by the Red MSX Source (RMS) survey (Lumsden et
al. 2002) that utilized the 18\arcsec\ resolution mid-IR survey by
the MSX satellite (Price et al. 2001) and high resolution,
multi-wavelength follow-up to yield a large, well-selected sample of
MYSOs. This is fairly complete out to about 15\,kpc for O star
luminosities (Urquhart et al 2011; 2012), and the GLIMPSE survey can easily
push this down to the B star range (Robitaille et al. 2008). However,
the IR spectral energy distributions of MYSOs and $\uchii$s are very
similar and therefore radio continuum data are vital to distinguish
these two evolutionary phases (Figure~\ref{cornishfig}) (e.g.,
Urquhart et al. 2007). Hence, radio observations are crucial in
distinguishing the `radio-quiet' MYSOs from the `radio-loud'
$\uchii$~regions.  The large numbers of very red objects in the
GLIMPSE survey with colours of embedded YSOs makes radio follow-up of
each one impractical. The uniform coverage of the CORNISH survey makes
this automatic.

The RMS survey also found many $\uchii$~regions, but such IR selected
samples are biased against very embedded objects such as those found
in IR dark clouds (IRDCs) (Egan et al. 1998). Many examples of
methanol maser sources (Ellingsen 2006), extended green objects (EGOs)
(Cyganowski et al. 2008) and faint 24\,\mic\ MIPSGAL sources (Rathborne
et al. 2010) are showing up in IRDCs. Is there radio continuum
emission associated with these very early phases of massive star
formation? Cyganowski et al. (2011) performed deep radio continuum
observations of a sample of EGOs and found mostly weak or non-detected
sources consistent with jets or winds in the diagnostic plots of Hoare
\& Franco (2007). The rarity of known MYSOs with luminosities above
10$^{5}$\,L$_{\odot}$ raises the question of whether early O stars ever
go through a detectable mid-IR bright MYSO phase or whether they
manifest themselves firstly as $\uchii$s.

The clean selection function of the CORNISH survey combined with
dynamical models and population simulation will allow us to determine
the lifetime of this phase as a function of luminosity (e.g., Davies
et al. 2011). Together with similar work for MYSOs from the RMS survey
and earlier phases from the Hi-GAL and SCUBA2 surveys this will yield
the first comprehensive picture of the evolutionary scheme from high
mass star formation. Modelling the flux, size and shape distributions
of the $\uchii$~region population will test the dynamical models of 
$\uchii$~regions and their relative locations in the dense molecular cloud
material will enable statistical tests of triggering scenarios.  These
tests are only possible from uniform, wide-area surveys with
sufficient sensitivity and resolution.

In the high-mass star formation context, a spatial resolution of about
1\arcsec\ is needed in a complementary radio survey for several
reasons. Firstly, the IR source density in regions of massive star
formation is very high in dense star forming regions. Hence, to
accurately identify the IR counterpart to a compact radio source,
arcsecond resolution is needed. Secondly, $\uchii$s themselves often
appear in complexes with separations of a few arcseconds (e.g., Sewilo
et al. 2004). Finally, as previous surveys have shown, to reveal the
structure of most $\uchii$s, a resolution finer than an arcsecond is
needed.

\subsection{Planetary and Proto-planetary Nebulae}

Zijlstra \& Pottasch (1991) estimated that there should be about
23\,000 PN in the galaxy, although only just over 3000 have been 
identified (Frew \& Parker 2010). Combined radio and IRAS searches
have been successful in turning up new PN (Van de Steene \& Pottasch 1995;
Condon, Kaplan, Terzian 1999).  A high resolution radio survey is
particularly suited to discovering young, compact PN that are heavily
reddened by line-of-sight extinction in the Galactic plane and we
expect about a thousand of these in the survey area. Their radio
flux densities are expected to be in the range of $5-50$\,mJy for typical
distances with sizes of a few arcseconds.  A complete sample over a
large region of the Galactic plane will constrain the density of PN
and hence their formation rate and distribution in the Galaxy. This
has implications for models of post-AGB evolution and mass-loss which
are not well constrained (e.g., Casassus \& Roche 2001).

As in the case of $\uchii$s, the immediate progenitors of PN - the
proto-PN (PPN) are also radio quiet since the central star has not yet
reached 30\,000\,K.  The radio-quiet PPN will also have very similar
mid-IR colours to the radio-loud PN and so the only way to distinguish
them is through their lack of radio emission. This is another
illustration of the usefulness of non-detections as well as detections
in the radio survey.  PPN hold the key to understanding the envelope
ejection process and testing the interacting winds models for the
shaping of PNe (e.g., Mellema \& Frank 1995).  It is not known whether
aspherical mass-loss at the end of the AGB is due to AGB star spin-up,
detached or common-envelope binaries (Mastrodemos \& Morris 1999) or
dynamically important magnetic fields (Gardiner \& Frank 2001). If we
take the birth rate of PN to be about 1 per year (Kwok 2000) and the
same for PPN with a PPN lifetime of 1500 years then we expect about
250 PPN in our survey region.

\begin{figure}
\centering
\includegraphics[width=8.5cm]{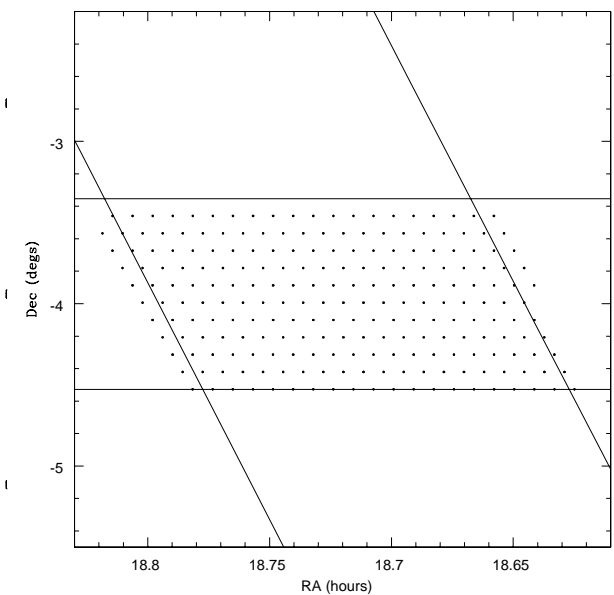}
\caption{Illustration of the pointing pattern for a typical 8\,hr
observing block. Each dot represents a pointing centre. Scans in RA
started at the bottom of the block and moved upwards before repeating
to improve the {\it uv} coverage. The diagonal lines represent $b=-1$
and $b=+1$ whilst the horizontal lines indicate the borders of this
block.  Note that pointings extended up to 6.4\arcmin\ outside of the
nominal survey area to ensure uniform sensitivity within it.
\label{block}}
\end{figure}

\begin{figure}
\centering
\includegraphics[width=8.5cm]{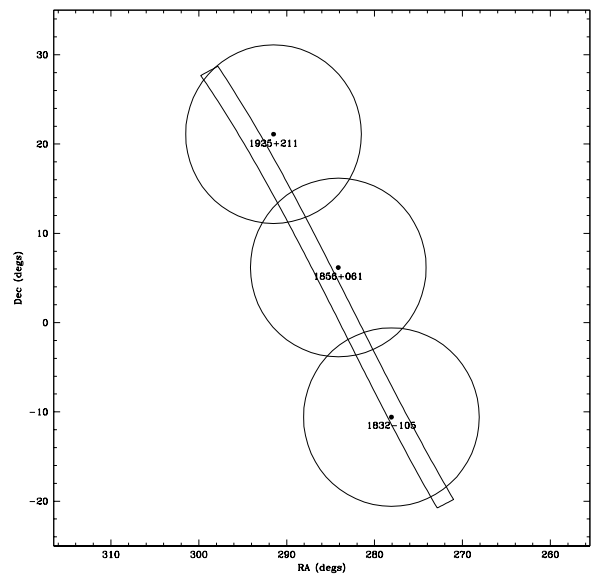}
\caption{The entire survey area with the locations of the three phase
  calibrators indicated. The circles have radii of 10\degs\ showing that
these calibrators were sufficient to cover the entire survey area.
\label{calibs}}
\end{figure}

\subsection{Evolved OB stars}

Mass-loss from the later evolutionary stages of massive stars (eg. OB
supergiants, Wolf-Rayet stars, and Luminous Blue Variables) is of
great importance in the evolution of the ISM in our own and other
galaxies, through both the kinetic power deposited by their stellar
winds and their strong ionizing radiation fields. Thermal radio
emission plays a key role in determining mass-loss rates for the
evolved massive stars since it arises at large distances from the
underlying stars where the outflow has attained a constant velocity
(e.g., Wright \& Barlow, 1975). Renewed interest in this important
input to the ISM has arisen due to clumpiness of the winds (Fullerton
et al. 2006). Radio deduced mass-loss rates may also provide evidence
of the bi-stability jump, a drop in ratio of wind-terminal velocity
and the escape velocity (Lamers, Snow \& Lindholm 1995), and perhaps
important to understanding of the LBV phase of stellar evolution (Vink
\& de Koter 2002). 

Many of these massive stars exhibit non-thermal emission in addition
to the thermal emission from their stellar winds (Bieging et
al. 1989), shown to be associated with massive binary systems, with
the emission arising where the stellar winds of the two companions
collides. The archetype of these colliding-wind binaries is WR140,
where the non-thermal emission has led to the determination of all the
orbit parameters, the distance and the wind-momentum and mass ratios
(e.g. Dougherty et al. 2005; Dougherty et al. 2011). Among WR stars
that exhibit non-thermal emission, over 90\,\% are binaries (Dougherty
\& Williams, 2000), and there is now evidence that a similar fraction
of non-thermal emitters is also observed in O-star binaries (de Becker
2007). Challenges remain for the colliding-wind
model. Not all stars with non-thermal emission are found to be
binaries, and many binaries exhibit thermal emission, which among
short-period systems might be attributed to absorption of the
synchrotron power. Alternative mechanisms for accelerating electrons
to necessary relativistic speeds have been proposed e.g. shocks within
the stellar winds, that work in single star scenarios.
 
Studies of both the thermal and non-thermal emitters are limited by
the small sample size of the detected systems, and the samples are
often biased via spectral type, optical luminosity, or known binary
systems. A large, unbiased sample of radio detected evolved OB stars
has previously not been available. Since they inhabit the densest
parts of the plane, there are likely many evolved OB stars hidden from
optical view by high extinction. For a
5\,GHz detection limit of 2\,mJy, the limiting radio luminosity is
$2.4\times10^{18}(D/{\rm kpc})^2$\,erg\,s$^{-1}$\,Hz$^{-1}$.  Using the 
empirical relationship between \mdot and $L_{\rm bol}$ from Garmany
\& Conti (1984) and the relationship between $L_{\rm 6cm}$ and \mdot
from radio flux from Bieging et al. (1989) we expect this survey
to detect all stars with a bolometric luminosity $L_{\rm bol} \ge
2\times10^{6}$L$_\sun$\ within $\sim1$\,kpc.

\subsection{Active Stars}

Sensitive radio observations have revealed the ubiquity of high energy
processes across the radio H-R diagram (G\"{u}del 2002). As well as
dynamo effects in convective envelopes, high-resolution radio
observations have revealed large-scale magnetospheric structures in
various stellar types, including both low- and intermediate-mass
pre-main sequence stars. Radio observations reveal dozens of such
objects in Orion, and Chandra now finds over 1000 active stars in
Orion (Garmire et al 2000).  With 5\,GHz flux densities reaching
several tens of mJy, these sources will be detected to several kpc.
Particularly intriguing is the role of magnetic fields in
intermediate-mass pre-main sequence stars, which are not expected to have
convective envelopes and hence large scale dynamos
(e.g., Garrington et al 2002). Combined radio, IR and X-ray studies will reveal how
common these magnetic phenomena are.

\begin{figure*}
\centering
\includegraphics[width=17.5cm]{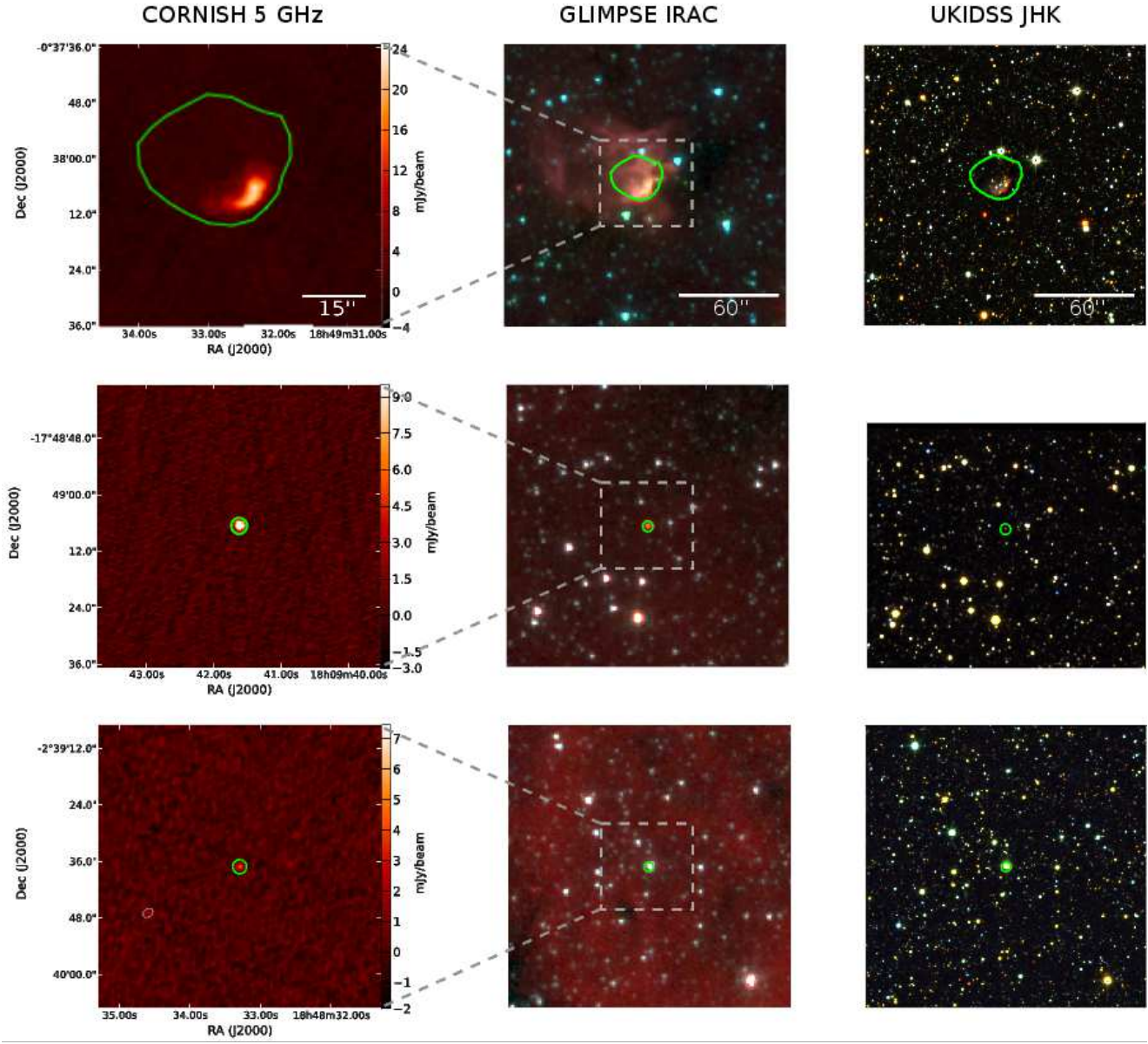}
\caption{Example data from the CORNISH survey alongside three-colour
  images from the {\it Spitzer} GLIMPSE (3.6\,\mic, 4.5\,\mic\ and
  8.0\,\mic) and UKIDSS (JHK) surveys. {\bf Top:} A typical cometary
  $\hii$~region, G032.1502+00.1329. Note the good correspondance
  between the radio emission and the brightest 8\,\mic\ emission in
  the GLIMPSE image. There is also strong extinction ahead of the
  cometary as expected if the OB star formed in a density
  gradient. {\bf Middle:} A candidate PN that was previously unknown,
  G012.3830+00.7990. It is clearly seen as an isolated source at
  8\,\mic\ with a very faint counterpart in the near-IR.  In the
  longer wavelength {\it Spitzer} MIPSGAL data (not shown here) it is
  bright at 24\,\mic\ but fainter at 70\,\mic\ unlike $\hii$~regions
  that are brighter at 70\,\mic. {\bf Bottom:} A
  previously unknown radio star, G030.2357$-$00.5719, that has blue
  colours in the near- and mid-IR.  \label{examples1}}
\end{figure*}

\begin{figure*}
\centering
\includegraphics[width=17.5cm]{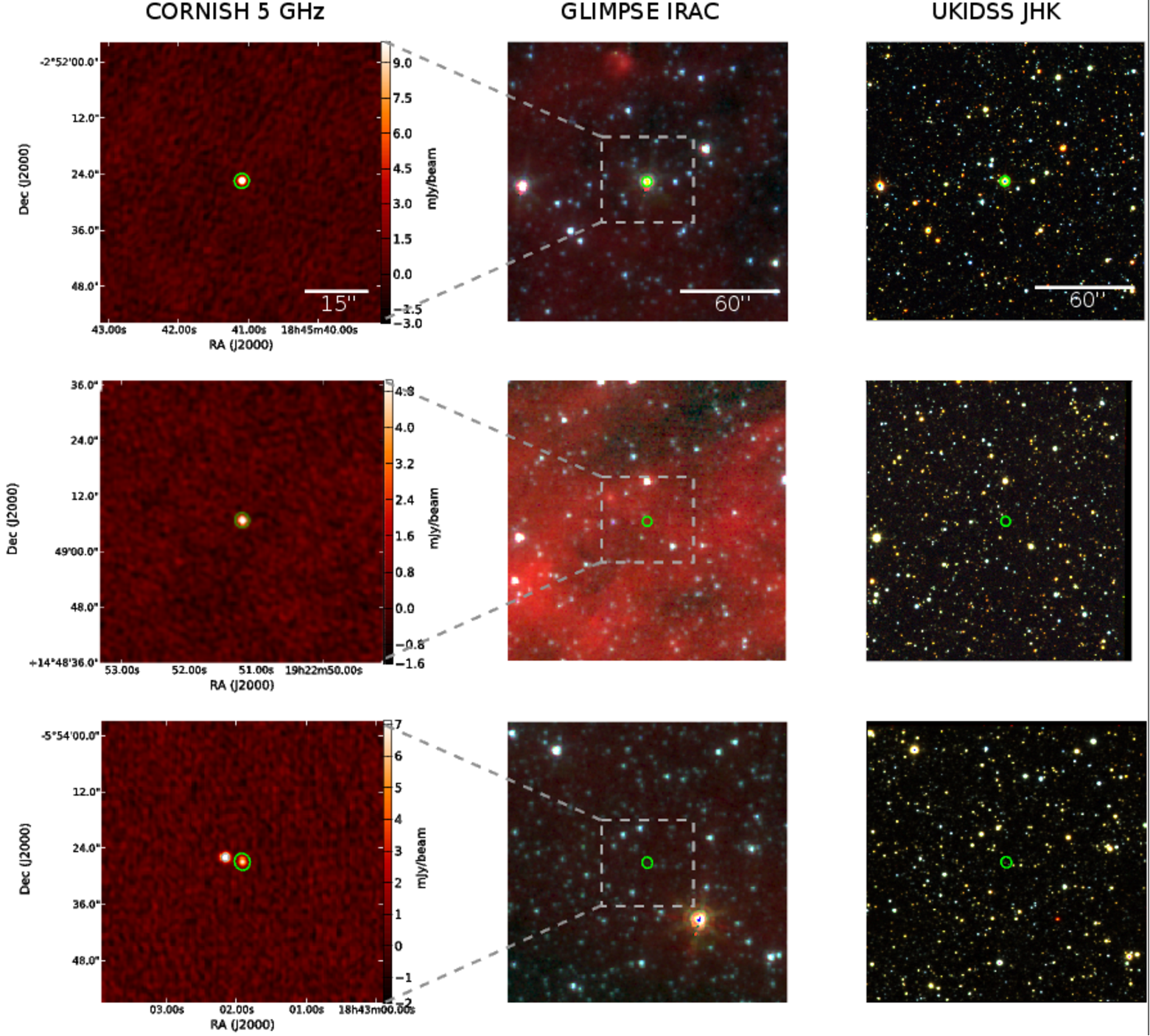}
\caption{Further example data.
{\bf Top:} A previously unknown radio star, G029.7188$-$00.0316, that
has red colours in the near- and mid-IR. 
{\bf Middle:} A candidate radio galaxy where we only see one unresolved
component, G049.6617$-$00.0543. No counterpart is seen in the IR. 
{\bf Bottom:} A candidate radio galaxy where we see two unresolved
components, G026.7179$-$00.8290. Again, no counterpart is seen in the
IR.  \label{examples2}}
\end{figure*}

\subsection{Active Binaries}

Mass transfer in close binaries often leads to strong activity in
radio and other wave-bands. RS CVn stars, cataclysmic variables and
others are likely to be detected in the survey. However, the most
important class of active binary are the so-called micro-quasars where
the accreting object is a black-hole or neutron star (Mirabel \&
Rodr\'{i}guez 1999; Fender 2002). These produce relativistic jets
analogous to AGN where ejections of plasma can be followed in real
time (Fender et al. 1999). These objects are perhaps the best laboratories for
studying the relativistic jet formation process and test magneto-hydrodynamic
models (e.g., Meier, Koide \& Uchida 2001).
The key signature of these jets is strong synchrotron emission
in the radio band and, since they often originate from massive stars, one
expects to find them close to the Galactic plane and in the locale of
giant molecular clouds.  In such dense environments the X-ray and
optical signatures of accretion may be heavily absorbed,
preventing the detection of accretion by more conventional
means. Hence, a co-ordinated radio and IR survey should uncover many
new examples of this phenomena.

There is a strong relationship between the X-ray state and radio
behaviour in these objects, which has led to the concept of disk-jet
coupling. Several black hole candidate objects have flat or inverted
radio spectra when in the low/hard X-ray state, indicative of the
presence of a self-absorbed compact jet supported by a hot X-ray corona.
The jets have been resolved by VLBI in GRS1915+105 (Dhawan et al. 2000)
and Cygnus X-1 (Stirling et al 2001). When the accretion rate increases
the thermal disk becomes dominant and the X-rays become softer. Radio
outbursts often occur during the transition between hard and soft states
(Fender, Homan and Belloni 2009) due to the formation of a strong shock
in the out-flowing jet, and radio interferometer observations are able
to follow the evolution of the flow (e.g., Fender et al 1999). Our survey
has the resolution to resolve such structures and at high accretion
rates the objects are detectable throughout the disk of the galaxy.

\subsection{Extragalactic Sources}

There will be a significant number of background extragalactic sources
detected during the survey. Using equation A2 in Anglada et
al. (1998), which is based on the source count data from Condon
(1984), at our completeness level we expect to detect about 2500
extragalactic sources in the 110 square degree area.  The vast
majority of these radio galaxies will be at too high a redshift to
determine distances via H I detections with current telescopes
(Magliocchetti et al. 2000; Obreschkow et al. 2009). As well as
penetrating the deepest part of the zone of avoidance these sources
will be useful in other respects. The most compact ones are potential
phase calibrators for use in future interferometric studies in the
Galactic plane.  Being at 5\,GHz the CORNISH survey will
preferentially pick up flat spectrum sources, which are potential mm
interferometric phase calibrators.  Extragalactic sources that are
also visible in the optical or near-IR can be used to bring together
the radio and optical coordinate reference frames which is important
for multi-wavelength high resolution studies in the Galactic plane.

Radio surveys can also deliver a set of extragalactic background
sources that can act as pencil beam probes of molecular clouds in
follow-up absorption line studies (Greaves \& Nyman
1996). Sufficiently strong sources could be used for Zeeman absorption
line measurements of the magnetic field strength (Crutcher 1999). For
instance, radio sources that lie behind dense, starless molecular
cloud cores can be used to probe the initial conditions for star
formation in these dense clouds with future sensitive instruments such
as the SKA. Polarized background sources can also be used to probe the
magnetic field of the Galaxy, individual $\hii$~regions and supernova
remnants via Faraday rotation (Taylor et al. 2009). High frequency
data are useful in disentangling components in sources with complex
Faraday depth spectra (O'Sullivan et al. 2012).

\section{Survey Specification and Implementation}

The scientific motivation described in the preceding section required
a radio continuum survey at around 1\,mJy sensitivity,
1\arcsec\ resolution and at a frequency significantly higher than the
1.4\,GHz traditionally used for surveys. Given the positive spectral
index of many of the target sources then the higher the frequency the
better. For VLA receivers the 8\,GHz band would normally be the optimum
choice in terms of sensitivity for objects with spectral index around
+0.6. However, due to the primary beam size, the time required to
survey a given area increases as $\nu^2$. Thus a
large survey is much more practical at 5\,GHz. The VLA 5\,GHz sensitivity is
similar to that at 8\,GHz, but the survey only takes half the time.
For a frequency of 5\,GHz the VLA in B configuration delivers a
resolution of 1.5\arcsec\ that satisfies the scientific requirement. A
significant fraction of the area to be surveyed is below declination
$-15\degs$ where it is necessary to switch to the BnA configuration to
ensure that the beam elongation does not degrade the resolution. As
with any single configuration snapshot survey the spatial dynamic
range is limited and structures larger than about 12\arcsec\ in size
are not imaged well in the CORNISH survey.

In order to image the full (8.9\arcmin) primary beam without image
degradation due to bandwidth smearing these observations
were carried out in spectral line mode. A 25\,MHz bandwidth in 4IF mode
was used to maximise the sensitivity. This has eight 3.1\,MHz channels
which degrades the peak response by only a few percent at the edge of
the primary beam.

\begin{figure*}
\centering
\includegraphics[width=16.0cm]{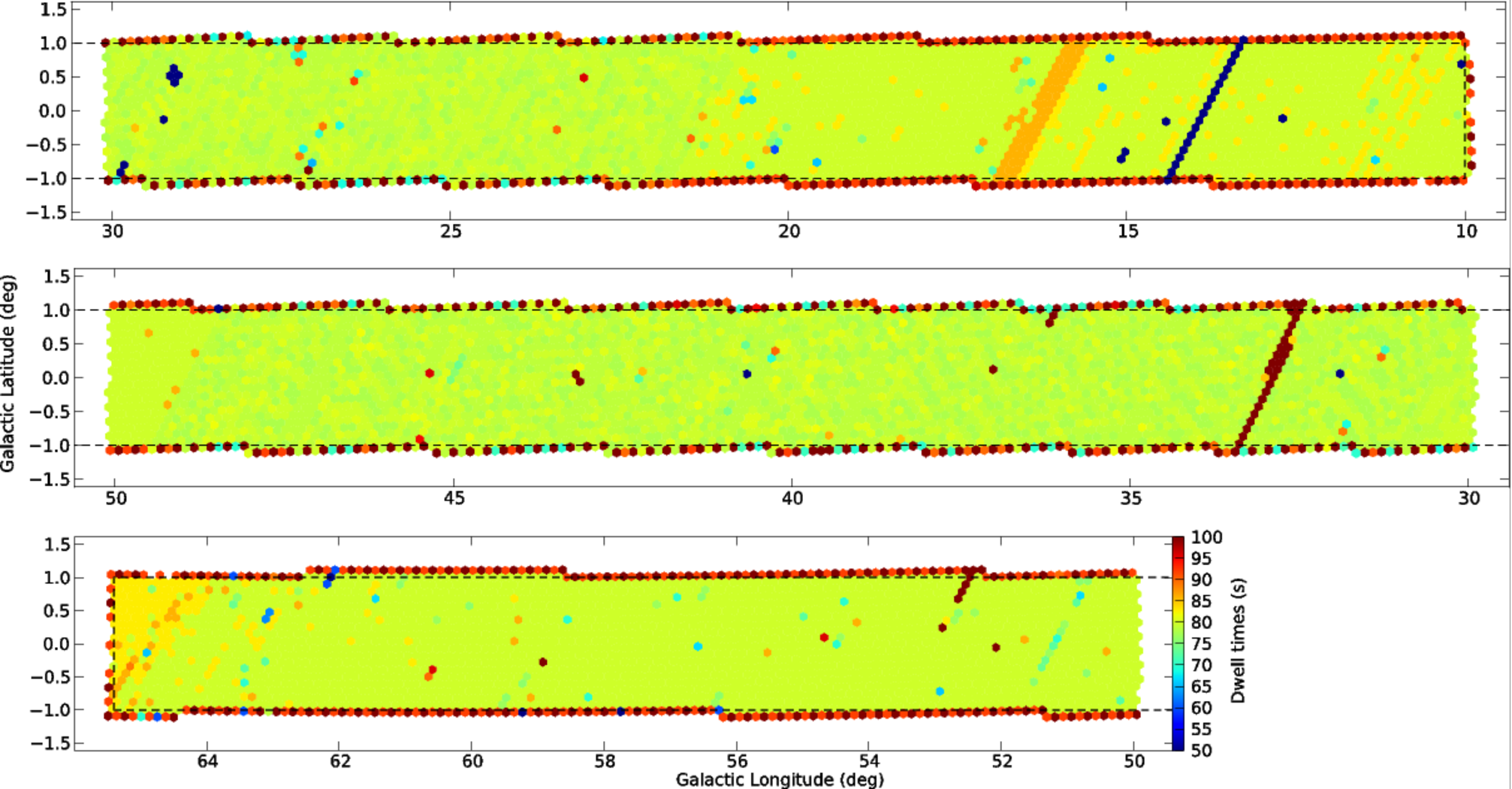}
\caption{Plot showing the dwell time achieved and the final area
  covered by the CORNISH survey.
\label{dwell}} 
\end{figure*}

\begin{figure*}
\centering
\includegraphics[width=16.cm]{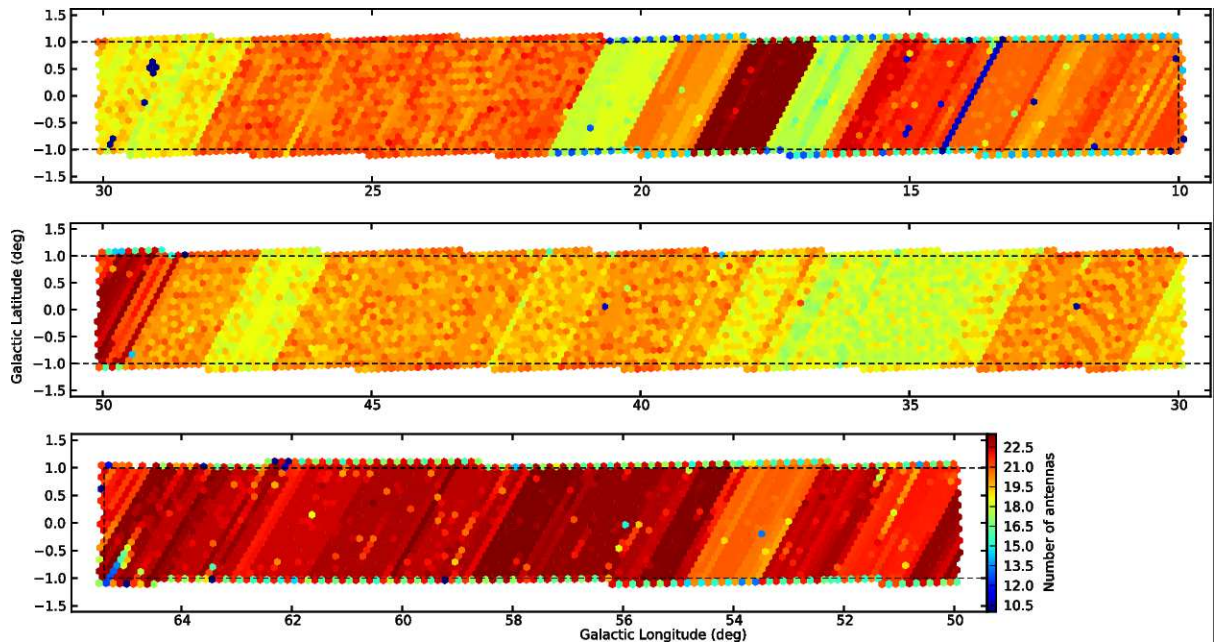}
\caption{Plot showing the effective number of antennas achieved during the
  CORNISH survey after flagging.
\label{nants}}
\end{figure*}

To achieve a noise level of 0.3\,mJy\,beam$^{-1}$ we required a total
on-source integration time of 90 seconds per pointing.  This noise
level is similar to that used by Wood \& Churchwell (1989) and Kurtz
et al. (1994) to detect and map $\uchii$ regions throughout the
Galaxy. If we use equations 1 and 3 from Kurtz et al. for $\uchii$s
with an unresolved flux density of 2\,mJy placed 20\,kpc away we find
that we can detect all objects powered by a B0.5V star or earlier.  

\begin{figure*}
\centering
\includegraphics[width=16cm]{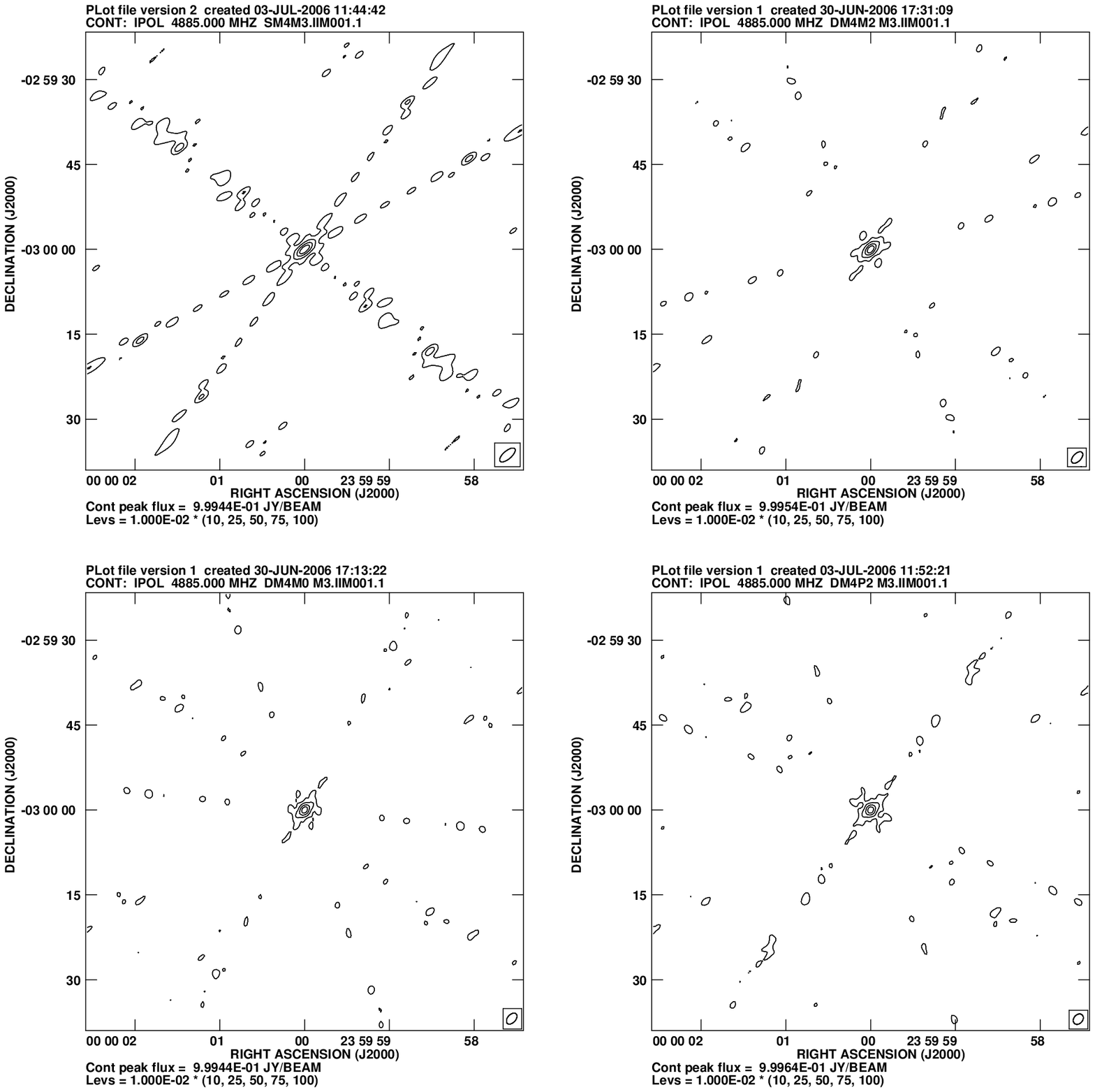}
\caption{{\bf Top left:} Simulated VLA beam for a single 2-minute
  snapshot at an hour angle of $-4$ hours and declination
  $-3\degs$. {\bf Top right:} Simulated VLA beam for two 1-minute
  snapshots at hour angles of $-4$~hours and $-2$~hours,
  i.e. separated by 2 hours. {\bf Bottom left:} Simulated VLA beam for
  two 1~minute snapshots at hour angles of $-4$~hours and 0~hours,
  i.e., separated by 4~hours. {\bf Bottom right:} Simulated VLA beam
  for two 1~minute snapshots at hour angles of $-4$~hours and
  +2~hours, i.e., separated by 6~hours.\label{beamfig}}
\end{figure*}

To achieve a sensitivity that is uniform across the survey to within
10\,\%, we adopted the NVSS hexagonal pattern of pointing centres
(Condon et al 1998).  Scaling the NVSS pointing pattern to 5\,GHz
gives a hexagonal grid separation $\Delta$ of 7.4\arcmin.  The
observations were carried out in the pseudo-holography mode developed
for the NVSS survey (see Condon et al. 1998), scanning in RA between
pointings. The separation between pointings in RA was
7.4\arcmin\,=\,(2~sin\,30\degs)$\Delta$ whilst those in Dec were
separated by 6.4\arcmin\,=\,(sin\,60\degs)$\Delta$. The pointing
pattern for a typical observing block is shown in Figure
\ref{block}. Forty such blocks were completed in 360 hours of time to
complete the survey. The survey was spread over two trimesters, with
$21\degs<l<49\degs$ being observed in the third quarter of 2006 and
the rest in the last quarter of 2007 and begining of 2008.

Three phase calibrators (1832$-$105, 1856+061 and 1925+211) were used to
cover the whole survey with all blocks being within about 10\degs\ of
one of these (see Figure~\ref{calibs}).  This continuity of phase
calibrator allowed additional 
cross checks of the overall calibration of the survey. A primary flux
calibrator (1331+305) was observed at the beginning and a secondary
flux calibrator (0137+331) at the end of each 8~hour observing run.
For the BnA observations 9 rows were completed in shorter 6 hour runs. 

Complete coverage of the survey area was achieved and indeed extended
to $l=65.4\degs$ that was the eventual limit of the GLIMPSE survey.
A log of the observation dates can be found in Table~\ref{obslog}.
Example images, representative of the different types of science
discussed in section~\ref{science}, are presented in
Figures~\ref{examples1} and~\ref{examples2}, alongside those from the
GLIMPSE  survey. The complementarity of the similar resolution radio
and IR datasets is very apparent.

\begin{deluxetable}{cc}
\tablecaption{Log of observations. \label{obslog}}
\tablewidth{0pt}
\tablehead{
\colhead{Dates} & \colhead{$l$ ranges observed}}
\startdata
12/07/2006 to 16/09/2006 & 21.1\degs~to 48.9\degs \\
28/09/2007 to 06/10/2007 & 10.0\degs~to 16.1\degs \\
27/10/2007 to 04/02/2008 & 16.1\degs~to 21.1\degs~and 48.9\degs~to 65.4\degs  \\
\enddata
\end{deluxetable}

Figure~\ref{dwell} shows the dwell time across the survey area where
the uniform coverage is clearly seen.  About 5\,\% of the survey
pointings were repeated due to weather or technical problems. Only one
scan had a single 45 second pass rather than the usual two (see below)
because the repeat scan also failed.  Along the edges at
$b=-1\degs$ and 1\degs on the end points of scans the on-source time
varied due to the nature of the raster scanning technique. During the
survey the VLA antennas were being retro-fitted into EVLA antennas,
which meant that there were often significantly less than the nominal
27 antennas. In addition to the usual flagging of antennas and
baselines due to technical issues this meant that the effective number
of antennas varied across the survey as shown in Figure~\ref{nants}.

\begin{figure*}
\centering
\includegraphics[width=16cm]{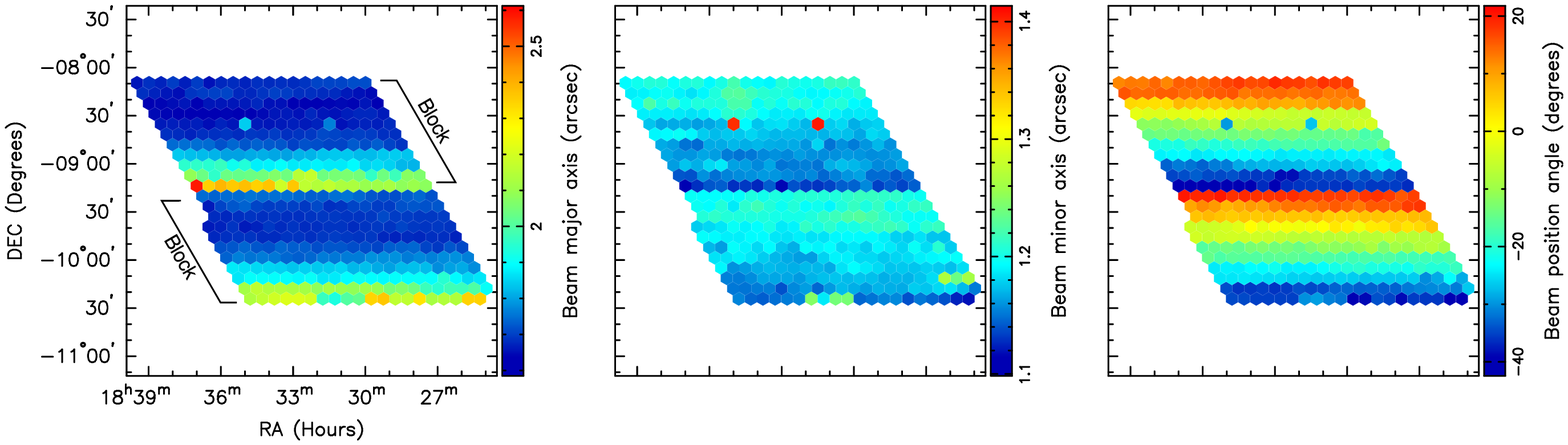}
\caption{The variation in the synthesised beam major axis, minor axis
  and position angle over two adjacent observing blocks covering
  $21.5\degs<l<24\degs$.} 
  \label{fig:block_beam_params}
\end{figure*}

\begin{figure*}
\centering
\includegraphics[width=16cm]{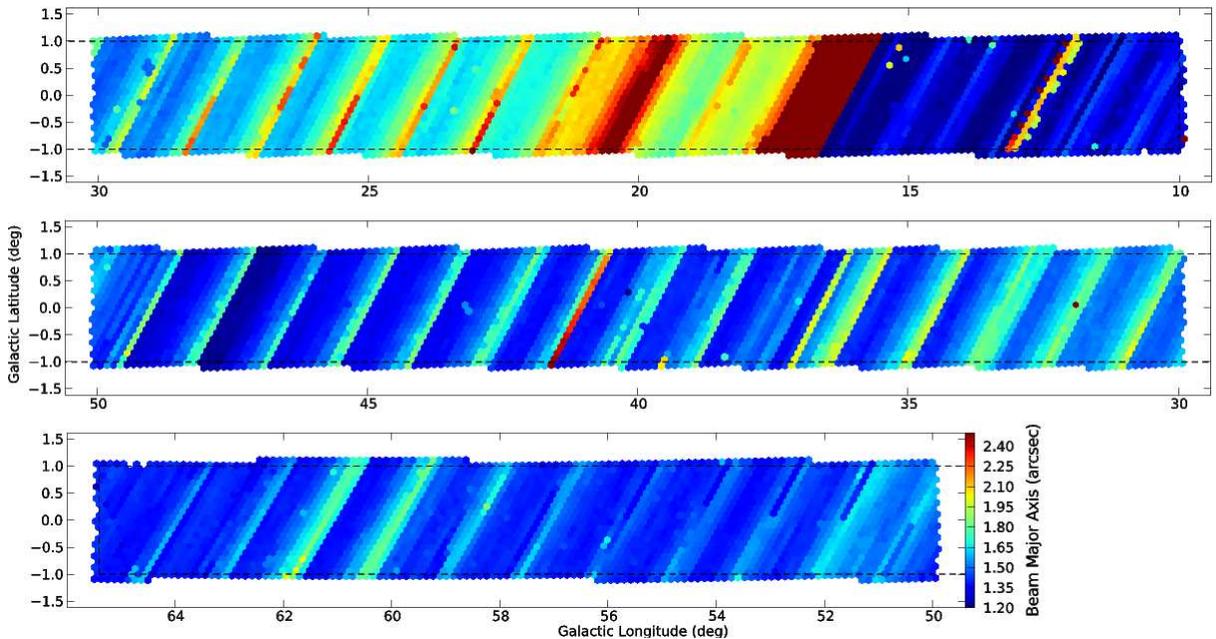}
\caption{Plot showing the major beam axis achieved throughout the
  survey.
\label{fullbmaj}}
\end{figure*}

The median noise level achieved over the survey area was
0.33\,mJy\,beam$^{-1}$ (Purcell et al. 2012) very close to that
expected. To determine the 
nature of the sources in the CORNISH survey it will be very useful to
evaluate the spectral index, which requires data at other radio
wavelengths. The principal existing source of these data will be the
1.4\,GHz plane surveys. For the survey carried out using the VLA in B
configuration by White et al. (2005) that overlaps with 91\,\% of the
CORNISH survey area their quoted median noise level is
0.90\,mJy\,beam$^{-1}$, although there is significant variation over
the area. This means that typically a source would need to have a
spectral index steeper than $-0.8$ to be guaranteed to be detected in
both. If we take the quoted 
completeness limit of 13.8\,mJy from White et al. (2005) then optically
thin $\hii$~regions with $\alpha=-0.1$ will need to be stronger than 12
mJy in CORNISH to be also detected in the 1.4\,GHz survey. For
optically thick $\hii$~regions with $\alpha=2.0$ this rises to 150\,mJy.
For 55\,\% of the CORNISH area multi-configuration VLA 1.4\,GHz data
exists from the MAGPIS survey (Helfand et al. 2006) with a noise level
that is more uniform and three times lower than the White et
al. (2005) survey. Therefore, for most non-thermal sources in
the CORNISH survey with a typical $\alpha=-0.6$ then sufficient data
exists to characterise their spectral index. However, for thermal
sources with a positive spectral index follow-up observations will be
required to measure the spectral index.

\begin{deluxetable}{rrccc}
\tablecaption{Simulated VLA beam parameters for the different
  snapshot strategies shown in Figure~\ref{beamfig}. \label{beams}}
\tablewidth{0pt}
\tablehead{
\colhead{HA$_{1}$ (hrs)}  & \colhead{HA$_{2}$ (hrs)} &
\colhead{$b_{\rm maj}$ (\arcsec)} & \colhead{$b_{\rm min}$ (\arcsec)} & \colhead{$b_{\rm maj}/b_{\rm min}$}}
\startdata
-4 & $-$ & 3.3 & 1.4 & 2.3 \\
-4 & $-$2  & 2.5 & 1.4 & 1.8 \\
-4 & 0   & 2.3 & 1.4 & 1.6 \\
-4 & +2  & 2.0 & 1.5 & 1.3 \\
\enddata
\end{deluxetable}

\begin{figure*}
\centering
\includegraphics[width=16cm]{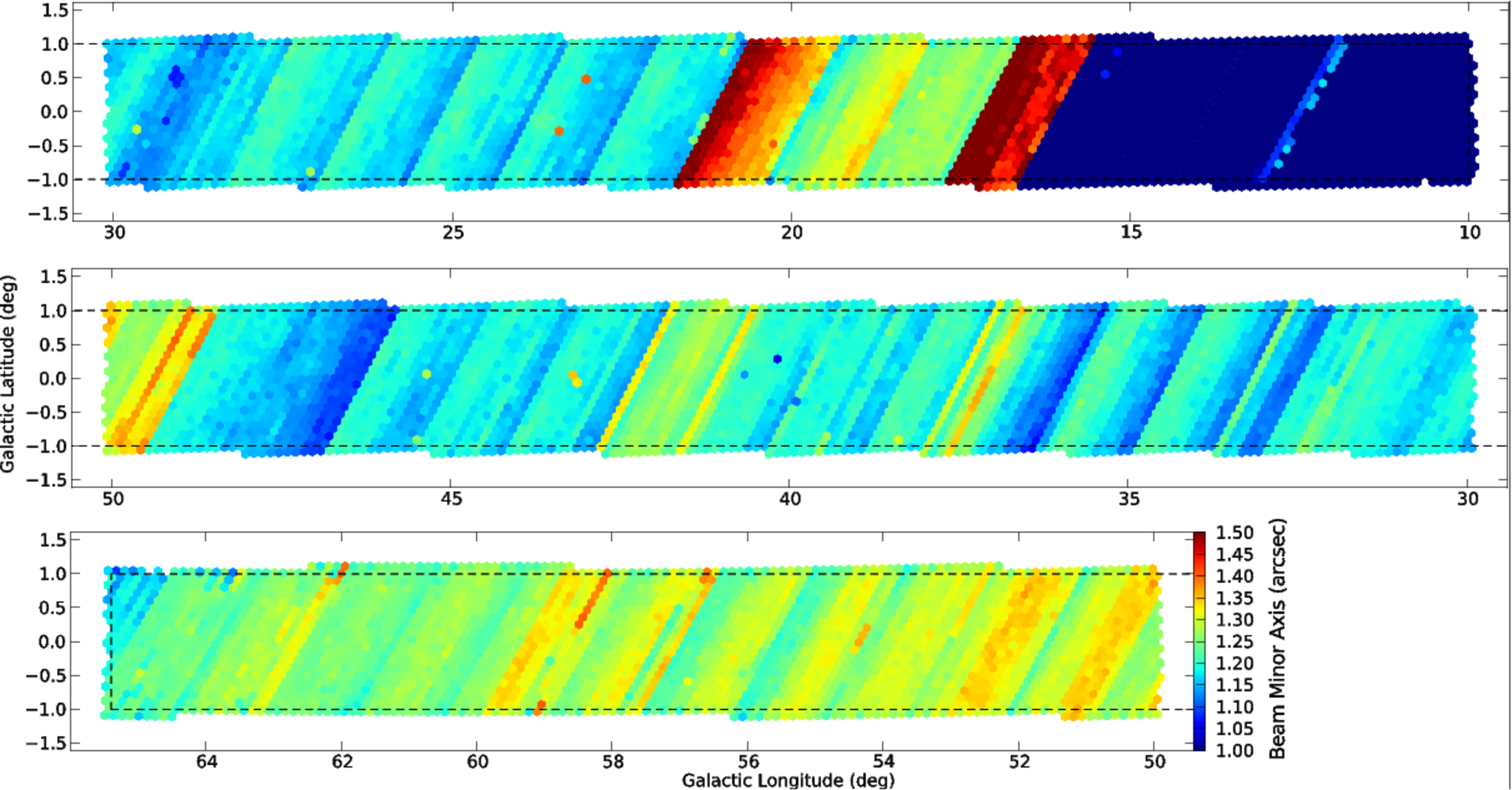}
\caption{Plot showing the minor beam axis achieved throughout the
  survey.
\label{fullbmin}}
\end{figure*}

\begin{figure*}
\centering
\includegraphics[width=16cm]{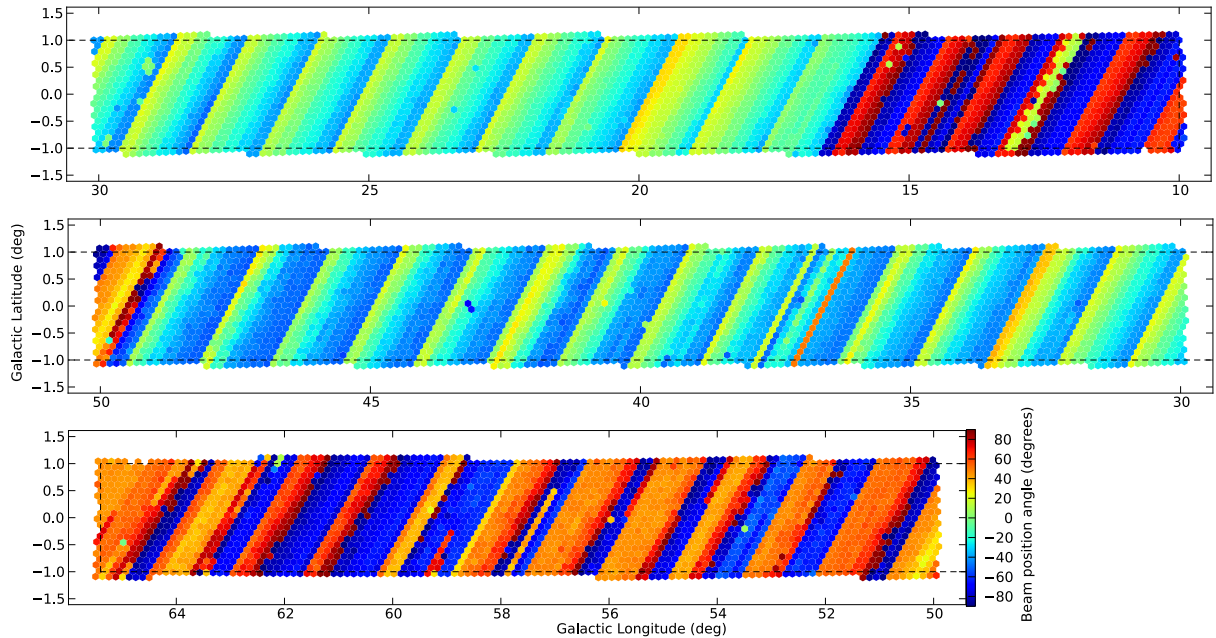}
\caption{Plot showing the beam position angle achieved throughout the survey.
\label{fullpa}}
\end{figure*}

\section{Beam Shape}

To improve the {\it uv} coverage and reduce the sidelobes the 90
second integrations were split into two separate snapshots of 45
seconds, each separated by 4 hours in hour angle. This strategy was
arrived at after simulations were made of the beam shape for various
hour angle separations.  Figure~\ref{beamfig} shows the simulated
beams for a single 2 minute snapshot as well as two one-minute
snapshots separated by 2, 4 and 6 hours. Table~\ref{beams} lists the
beam parameters for each of these simulations. The two snapshots
clearly give better beam structure, with the ones separated by 6 hours
giving the best beam shape. However, we used a 4 hour separation for
the survey as this made implementation and scheduling easier in the 8
hour observing blocks when the Galactic plane was observable.

\begin{figure}
\centering
\includegraphics[width=8.5cm]{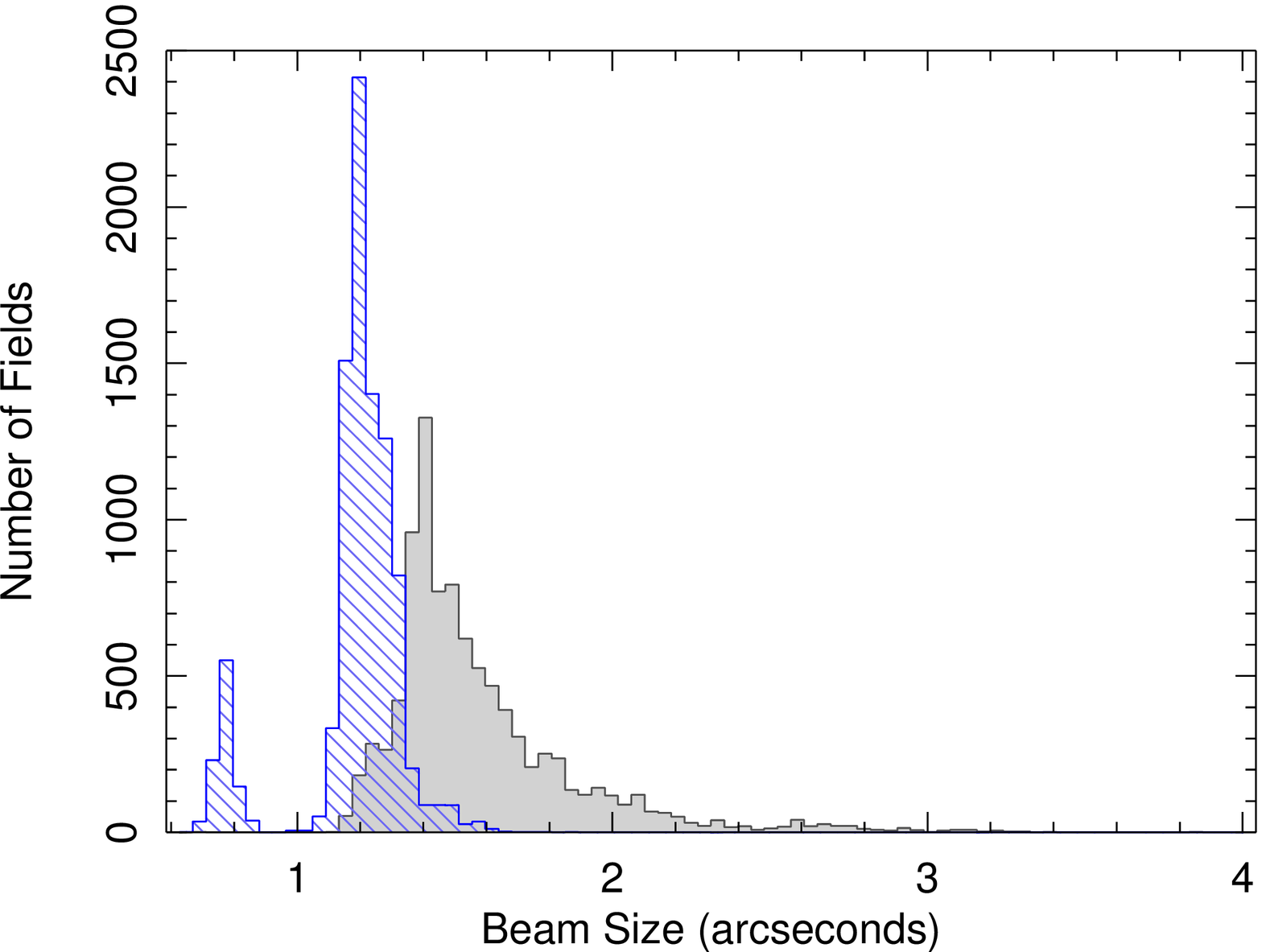}
\caption{Distribution of major (grey) and
    minor (left-slanted) beam axes.} 
  \label{fig:hist_beam_majmin}
\end{figure}

\begin{figure}
\centering
\includegraphics[width=8.5cm]{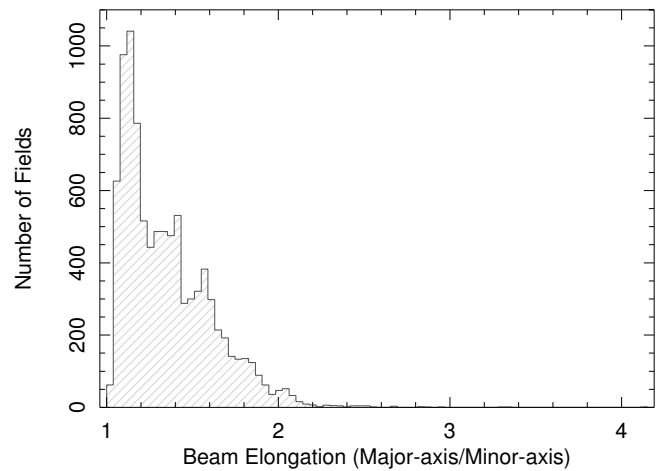}
\caption{Distribution of beam elongation.} 
  \label{fig:hist_beam_elong}
\end{figure}

A typical scanning row contained 20 pointings (see
Figure~\ref{block}), which including the 
drive time of about 5 seconds between pointings took about
17~minutes. Following each scan row a phase calibration of 2~minutes
on-source was carried out. Eleven of these rows formed an observing
block of about 3.6~hours, after which the whole block was repeated,
giving the second 
snapshot at the appropriate 4 hour separation in hour angle. For the
BnA observations the observing runs were only 6~hours long due to the
lower elevation and the two snapshots were then only 3~hours apart.  

The actual beam shape achieved in two consecutive blocks from survey
observations carried out during 2006 is shown in
Figure~\ref{fig:block_beam_params}. This is the synthesised beam
(imaged with a robust parameter of zero) after the combination of the two
individual 45 second snapshots.  The major axis of the beam is
slightly worse at the beginning of the blocks (low declination end)
than that predicted in Table~\ref{beams} for a similar
declination. This is because scheduling constraints meant that the
2006 blocks did not quite transit mid-way through the 8 hour observing
session which resulted in a more asymmetric beam than otherwise
achievable. The beam at the end of the block (high declination end)
should also have a major axis of about 2.3\arcsec\ from a combination
of the first snapshot at hour angle 0 and a second one at hour angle
+4. Another manifestation of the asymmetry in the scheduling is in the
beam position angle that rotated from $-40\degs$ to
+20\degs\ throughout the block.  The minor axis of the beam varied
less across the blocks, as expected from Table~\ref{beams}.

The beam parameters over the full survey area are shown in
Figures~\ref{fullbmaj} to~\ref{fullpa}. Observations conducted during
the 2007/2008 were scheduled more 
symmetrically than in 2006, although the inevitable
variation in beam shape is still present. The large change around
$l=16\degs$ and below is due to the switchover to the BnA
configuration for declinations below $-15\degs$.  Here the position
angle of the elongated beam changes from being mostly north-south to
an east-west direction. 

Histograms of the beam parameters achieved are shown in
Figures~\ref{fig:hist_beam_majmin} and~\ref{fig:hist_beam_elong}.  The
median major axis size achieved was $1.5\arcsec$ with a standard deviation of
$0.3\arcsec$ and a tail up to about 3\arcsec. As expected the minor axis
size has a fairly narrow distribution around 1.2\arcsec\ apart from
the small number of higher resolution observations due to the BnA
configuration data. The distribution of beam elongations shows that
98\,\% of fields have elongations less than 2 and 74\,\% less than 1.5 in
agreement with the predictions of Table~\ref{beams}. This shows that in the
main the adopted strategy of two snapshots separated by 4~hours in
hour angle did deliver the desired constraints on beam
shape.

\section{Conclusions}

The CORNISH survey is a new sensitive, high spatial resolution
Galactic plane survey at 5\,GHz. It is targeted at compact thermal
sources and $\uchii$ regions in particular.  Conceived to complement the
{\it Spitzer} GLIMPSE survey, it will systematically address key questions in
massive star formation such as the lifetime and evolution of the $\uchii$
phase as a function of exciting star parameters and
environment. Uniform coverage of the northern GLIMPSE region enables
the distinction between radio-loud and radio-quiet objects such as
MYSOs/$\uchii$s and PPN/PN that otherwise have very similar mid-IR
colours.  Legacy survey science in combination with other recent and
upcoming optical, near-IR, far-IR, sub-mm and longer wavelength radio
surveys of the plane will be possible with applications in evolved
stars, active stars and active binaries. 

Data taking for the CORNISH survey has been completed and the basic
goal of uniformity of coverage achieved. The data reduction, source
extraction and statistical properties of the survey and sources will
be discussed in the forthcoming paper by Purcell et al. (2012).  
At that time the catalogue and image data will be publically available
at http://www.ast.leeds.ac.uk/cornish. 

We have also achieved the aim of controlling the beam shape across the
survey such that the elongation of the synthesised beam is mostly less
than 1.5 and nearly always less than 2. This was achieved by using a
strategy of two snapshots separated by 4 hours in hour angle. A more
uniform and rounder beam could be achieved by more snapshots but that
would increase the overheads and has diminishing returns with the
VLA. Our simulations show that snapshots separated by 6~hours in hour
angle would have produced a better beam. However, this would have
complicated the scheduling of the survey into 8~hour runs and more
than likely compromised the very uniform coverage that we ultimately
achieved. Such tradeoffs will be the case for any survey with the EVLA
at these declinations and for future Galactic plane surveys in
particular.

\acknowledgments

The National Radio Astronomy Observatory is a facility of the National
Science Foundation operated under cooperative agreement by Associated
Universities, Inc.

\end{document}